\documentclass[runningheads]{llncs}
\usepackage{graphicx}
\usepackage{amsmath}
\usepackage{multirow}
\usepackage{cite}
\usepackage{comment}
\begin{document}
\title{Characterizing Continual Learning Scenarios for Tumor Classification in Histopathology Images
% \thanks{Supported by organization x.}
}
\titlerunning{Continual Learning for Tumor Classification in Histopathology Images}

\author{Veena Kaustaban\inst{1} \and Qinle Ba\inst{1} \and Ipshita Bhattacharya\inst{1}\thanks{Work done at Roche} \and Nahil Sobh\inst{1} \and Satarupa Mukherjee\inst{1} \and Jim Martin\inst{1}\and Mohammad Saleh Miri\inst{1} \and Christoph Guetter\inst{1} \and Amal Chaturvedi\inst{1}}
% 1{Kaustaban,Veena} 2{Qinle, Ba} 3{Bhattacharya,Ipshita} 4{Sobh, Nahil} 5{Mukherjee, Satarupa} 6{Martin, Jim} 7{Miri, Mohammad Saleh} 8{Guetter Christoph} 9{Chaturvedi, Amal}
\authorrunning{Kaustaban V et al.}

\institute{Roche Sequencing Solutions, Santa Clara, CA, USA
\email{qinle.ba@roche.com}}

\maketitle              % typeset the header of the contribution
\begin{abstract}
Recent years have seen great advancements in the development of deep-learning models for histopathology image analysis in digital pathology (DP) applications, evidenced by the increasingly common deployment of these models in both research and clinical settings. Although such models have shown unprecedented performance in solving fundamental computational tasks in DP applications, they suffer from catastrophic forgetting when adapted to unseen data with transfer learning. With an increasing need for deep-learning models to handle ever-changing data distributions, including evolving patient population and new diagnosis assays, continual learning (CL) models that alleviate model forgetting need to be introduced in DP-based analysis. However, to our best knowledge, there’s no systematic study of such models for DP-specific applications. Here, we propose CL scenarios in DP settings, where histopathology image data from different sources/distributions arrive sequentially, the knowledge of which is integrated into a single model without training all the data from scratch. We then established an augmented dataset for colorectal cancer H\&E classification to simulate shifts of image appearance and evaluated CL model performance in the proposed CL scenarios. We leveraged a breast tumor H\&E dataset along with the colorectal cancer to evaluate CL from different tumor types. In addition, we evaluated CL methods in an online few-shot setting under the constraints of annotation and computational resources. We revealed promising results of CL in DP applications, potentially paving the way for application of these methods in clinical practice.

%\keywords{Digital Pathology  \and Continual Learning \and Few-shot Online Continual Learning \and Tumor Tissue Classification.}
\end{abstract}

\section{Introduction}
Deep learning has achieved unprecedented performance in solving complex problems in digital pathology (DP) based analysis \cite{cruz2014automatic, jm16, s18, b17}. There are, however, two challenges in model development for DP applications. Firstly, the increasing volume of data waiting to be analyzed. The increasingly wide adoption of DP along with development of new assays with, for example, the emerging multiplexing technologies, produce a near continuous stream of data to analyze \cite{ardon2021digital, campanella2019clinical }. Secondly, evolving data distributions for modeling, such as varying image digitization conditions, diverse patient populations, onset of new diseases and so on. For example, it is commonly observed that the use of different staining chromogens, changes in stainers, digital scanners and vendor platforms all result in a shift in appearance of the digitized images. Such ever-evoling image data call for iterative model development and model update even after model deployment. 

A general strategy to learn from multiple datasets considers training all the existing and newly arrived data from scratch. Such an approach in general ensures performance, but requires  storage of all the data and an increasing amount of computing resources each time a model is updated, which is not only inefficient, but also increases development cost and delays algorithm deployment/update. On the other hand, updating a model with transfer learning, albeit with no extra cost of data storage and computing resources, is known to suffer from catastrophic forgetting \cite{kprvdrmqra17}. To efficiently and effectively adapt a model to new data streams without forgetting of learned knowledge, continual learning (CL) algorithms have been proposed and investigated in the machine learning community \cite{kprvdrmqra17, crre18, lr17, rksl17}. However, these CL algorithms have largely been tested on relatively less complex datasets from non-DP domains, including MNIST \cite{crre18, lr17, art18} and CIFAR \cite{crre18, lr17, rksl17,art18}, the insights from which thus cannot be directly transferred to DP data. In the biomedical imaging field, despite a number of studies for medical imaging \cite{lss20, phh21, bgk18, zhang2021comprehensive, yang2021continual, bayasi2021culprit},  to our best knowledge, there is no systematic study so far on CL performance on deep learning/machine learning models designed for histopathology images, and consequently, no appropriate CL baselines for DP. In this study, we compare different CL methods with a sequential stream of DP images. New streams of data were designed to simulate real world scenarios where shifts in distribution, class labels or data volume can occur and thus the need to update an existing model.

Our contributions are three-fold: (1) we identify continual learning scenarios practical for DP-based analysis and systematically characterized the performance of recent CL methods in these scenarios (Section 2); (2) we establish a dataset with augmented H\&E images, simulating data streams from multiple data sources (Fig. 1), including different scanners, stainers and reagents, to evaluate CL methods (Fig. 1 and Fig. 2); (3) we have explored the feasibility and provided insights into performance of both learning continuously from more than one type of tumor (Fig. 4) and applying online CL methods (Fig. 3). 

\section{Continual Learning Scenarios in DP}
  We identify four learning scenarios \cite{crre18, vt19, de2021continual} for DP-based analysis in the continual learning (CL) framework. \textbf{Data Incremental Scenario (Data-IL)} Continuously learn from new streams of data from an identical underlying distribution during model development or update. Data-IL does not consider new classes or considerable shifts of data distribution and is thus expected to be the easiest among all CL scenarios. In DP, it is common for expert pathologists to generate annotations for a dataset in multiple batches and thus each batch can be considered as a sample of the same data distribution. \textbf{Domain Incremental Scenario (Domain-IL)} Continuously update a model with new streams of data from shifted distributions. Examples in clinical settings include variations of tissue processing/staining reagents, patient populations, scanning instruments and so on. Note that in practice the extent of domain shift is a continuum and challenging to predict beforehand, and thus the categorization of a scenario into Data-IL or Domain-IL is a design choice and/or empirical decision. As an extreme example, multiple tissue types can be sequentially included into the same class label, where there are domain shifts within the same class. \textbf{Class Incremental Scenario (Class-IL)} Continuously extend an existing model to unseen classes with new streams of data. Since the knowledge to learn from each batch of unseen classes is largely non-overlapping, this scenario is expected to be the hardest among CL scenarios. For example, adding new tissue types in a tumor classification model or adding new cell types into a cell detection model for AI-assisted diagnosis. \textbf{Task Incremental Scenario (Task-IL)} Any of the above scenarios can be considered as Task-IL if each data stream is defined as a new task and during inference prior knowledge is always given as to which task (task identity, i.e. task ID) the test data should be predicted with. In this study, we tested the task IL scenario with new classes introduced at each task. For example, a clinical-grade model incorporates patient data from different populations and uses patient meta-data as prior (i.e. task ID) for generating prediction.

\section{Materials and Methods}
\subsection{Datasets}
\textbf{CRC:} We characterized CL model performance with a colorectal cancer (CRC) dataset from \cite{khm18}, whose training and test set are composed of stain-normalized \cite{mna09} 224x224 patches at 20x from hematoxylin and eosin (H\&E) stained whole-slide images (136 patients) with tile-wise class labels for 9 tissue types - tumor, stroma, normal, lymphocyte-rich, mucin, muscle, adipose, debris, background. To establish a class-balanced dataset, we randomly selected 8700 images from each class for training (7000 train; 2700 validation) and the entire test set (7150 images) for testing. All the selected images were used for stain augmentation (See Section 3.2). \textbf{PatchCam:} To characterize the effectiveness of CL methods on learning from multiple tissue types, PatchCam from \cite{b17} was used for breast cancer classification, which comprises 96x96 patches with class labels of “normal” and “tumor” from 400 breast cancer H\&E whole-slide images at 20x resolution. We stain-normalized \cite{mna09} the patches to be consist with colorectal cancer dataset.

\subsection{Methods}
\textbf{Simulating Domain Shifted Data Streams} To systematically evaluate the CL algorithms in DP settings, we need a sequence of data to simulate CL scenarios. 
\begin{figure}
\includegraphics[width=\textwidth]{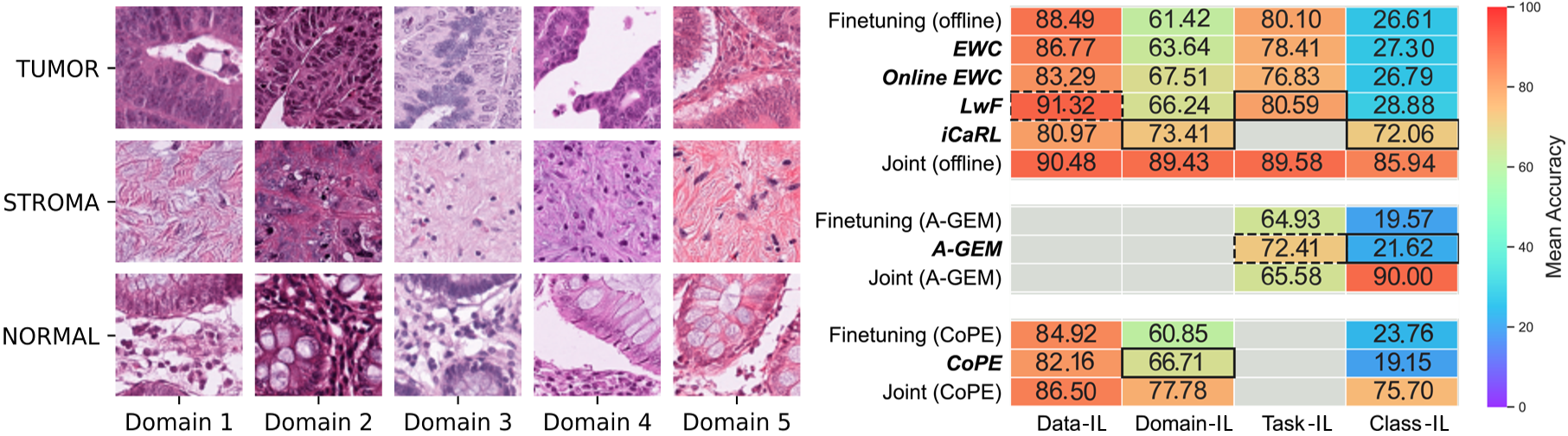}
\caption{Left: Example augmented CRC images from 5 domains (3 of 9 classes are shown). Right: Test accuracy of offline and online methods on augmented CRC. Online methods were compared to online baselines. Black outline: best performing method in the particular scenario exceeding finetuning (lower bound). Dashed outline: best performing method in the particular scenario exceeding joint training (upper bound).} \label{fig1}
\end{figure}
To the best of our knowledge, no such benchmark has been proposed for DP-based analysis. We thus generated an augmented CRC dataset comprised of multiple subsets, each with distinct colors and stain intensities, simulating the commonly observed variations of stain appearance (i.e. domain shifts) from multiple data sources. Except for the images provided by CRC (Domain 1), we simulated 4 additional data sources (Domain 2-5) (Fig. 1 Left). For each data split, We first randomly divided images in each of the 9 classes into 5 splits, kept 1 split unchanged for Domain 1 and performed stain unmixing \cite{ruifrok2001quantification} for the remaining splits. We further applied the following augmentation to remix the stain-unmixed intensity images into augmented H\&E and refer to this augmented dataset as \textbf{augmented CRC}. Pseudocode for creating this augmented dataset is at https://github.com/kaustabanv/miccai2022-cl-in-dp.
\begin{itemize}
\item Domain 2 (increased stain intensity, simulating, for example, concentration increase of the eosin and/or hemotoxylin solutions, each with a different extent of change): eosin intensity was randomly increased with a scaling factor sampled from [1.75, 2.75], hematoxylin from [1.5, 2.0]. 
\item Domain 3 (decreased eosin stain intensity, simulating, for example, slides prepared from many years ago with fading stain): eosin intensity scaling factor sampled from [0.4, 2.75].
\item Domain 4 (change of hue, simulating, for example, change of reagent manufacturer, scanner or stainer): eosin hue changed by a scaling factor sampled from [-0.05, -0.03], hematoxylin from [0.05,0.08]. 
\item Domain 5 (change of hue and saturation, simulating, for example, change of reagent manufacturer, scanner or stainer): eosin hue changed by [0.03, 0.05]; saturation increased for eosin by [1.2, 1.4] and hematoxylin by [1.1, 1.3].
\end{itemize}
\textbf{Qualitative Assessment of Augmented CRC} We presented four randomly selected images from each domain to two pathologists. Both of them confirmed that images of Domain 1-4 look realistic and one pathologist thought the Domain 5 is slightly deviating from majority of DP images from our organization.\\
\textbf{CL Setup with Augmented CRC} For \textit{Domain-IL}, the 5 domains were used as individual batches of data with domain shifts, where each domain contains examples from all 9 classes (e.g. for training, 8700/5 = 1740 examples per class per domain). In \textit{Class-IL}, each data batch contains all examples from the new classes and thus the same number of examples from each domain for each class. In \textit{Data-IL}, each data batch was randomly selected and roughly contained the same number of examples from each domain. \\
\textbf{CL Methods:} Recent proposed CL methods can be largely classified into three categories: replay, regularization and parameter isolation \cite{vfa21}. With replay methods, original images \cite{rksl17}, deep representations \cite{vt19} or model-generated pseudo samples \cite{slkk17} are selected via various heuristics, stored in memory and replayed at later learning stages to overcome forgetting. Regularization-based methods \cite{kprvdrmqra17, schwarz2018progress, zenke2017continual} avoid storing examples and instead include a regularization term in the loss function to penalize model updates that could lead to large deviation from an existing model, thus avoiding forgetting of learned knowledge. Parameter isolation methods assign different model parameters to each task. Note that while replay and regularization apply to all scenarios, parameter isolation only applies to Task-IL, because such methods are designed to either pre-allocate different parts of the network for different tasks \cite{asmk18}, or freezing previous model parameters \cite{rrdskkph16}: both require multi-headed designs and use task IDs to select a task-specific head during inference. Here we evaluate the following CL methods.
\begin{itemize}
\item\textit{\textbf{EWC/Online EWC:}} Regularization-based methods that use model parameters as prior \cite{kprvdrmqra17} to learn new data. The loss function includes a regularization term, which penalizes large changes to network weights that are important for previous tasks (knowledge from previous data) and is controlled by a weighting hyperparameter. The magnitude of penalty for each model weight is proportional to its importance for previous tasks. In EWC, the number of quadratic penalty terms grows linearly with the number of tasks. Online EWC \cite{schwarz2018progress} uses a single quadratic penalty term on parameters whose strength is determined by a running sum of parameter importance from previous tasks. Online EWC includes an additional hyperparameter to determine the magnitude of decay of each previous task’s contribution.
\item\textit{\textbf{LwF:}} Data-focused regularization method \cite{lh16} which distills knowledge from a previous model to an updated model trained on new data. The loss function has an additional distillation loss for replayed data. Each input is replayed with a soft target obtained using a copy of the model stored after finishing training on the most recent task.
\item\textit{\textbf{iCaRL:}} Replay-based method using stored original images from previous data streams \cite{rksl17}. This method stores a subset of most representative examples per class in memory (GPU, CPU, disk, etc), which are chosen according to approximate class means in the learned feature space. The memory size can be determined by grid search and/or availability of computing resources.
\item\textit{\textbf{A-GEM:}} In this online constrained replay-based method \cite{crre18}, model updates are constrained to prevent forgetting by projecting the estimated gradient on the direction determined by randomly selected samples from a replay buffer. Buffer size and sampling size from the buffer are both tunable hyperparameters. In the literature, A-GEM was tested only in Class-IL and Task-IL scenarios as reported in \cite{crre18}.
\item\textit{\textbf{CoPE:}} An online rehearsal based approach \cite{de2021continual}, which enables rapidly evolving prototypes with balanced memory and a loss function that updates the representations or pseudo prototypes (i.e. from model-generated pseudo samples), thus alleviating catastrophic forgetting. We tested CoPE in Data-IL, Domain-IL and Class-IL scenarios. 
\end{itemize}
\textbf{Model Evaluation Strategies} We refer to learning from a batch of data for CL as an experience. Within each experience, we split data into train, validation, test (see Section 3.1; e.g. images from the training set of augmented CRC were kept as training). We report (1) accuracy averaged over test sets at the end of training the last experience from multiple repeats (Fig. 1 Right, Table S1) and (2) accuracy over multiple repeats at each experience (Fig. 2-4). (3) While evaluation on test sets from past experiences indicates how well the model retains past knowledge without forgetting (backward transfer), accuracy on current/future test sets reflects new knowledge assimilation (forward transfer) (Fig. S1-S4). \\
\textbf{Baselines} All the methods were compared against two baselines: 1) joint training (upper bound), where a model was trained on all the data available so far; 2) finetuning (i.e transfer learning; lower bound), where a model was trained sequentially with exposure to only the data from the current experience. For A-GEM and CoPE we also compared the online versions of baselines for fair comparison: the A-GEM baselines were trained in an online setup for 1 epoch and for the CoPE baselines, the experiences were divided into mini experiences (128 samples each) in a data incremental fashion and trained for 1 epoch per mini experience, similar to its set-up in Data-IL, Domain-IL and Class-IL scenarios. 

\section{Experiments}
% \textbf{Implementation details}
\subsection{Implementation and Results} We used a ResNet-18 \cite{he2016deep} feature extraction layers followed by a single-layer nearest mean classifier \cite{rksl17}, where the feature vectors generated from the feature extractor were used to assign the class label with the nearest prototype vector. Training was conducted with 15 epochs and a batch size of 16 on 1 NVIDIA V100 or A100 GPU for 3-5 repeats. For Task-IL, in the multi-headed designs, each head was trained for a different task. Models were optimized with Stochastic Gradient Descent starting with a learning rate of 0.1 and momentum of 0.9 and a weight decay of 0.00001 applied after epochs 10 and 13. In all scenarios we ran grid search for best hyperparameters (See details at https://github.com/kaustabanv/miccai2022-cl-in-dp).\\ 
%\subsection{Results}
\textbf{Comparison of Offline CL Methods on Augmented CRC} Overall, the accuracy of joint baseline was comparable across scenarios, while the performance of finetuning varies among scenarios with best performance on Data-IL and worst on Class-IL (Fig. 1 Right and Table S1).
\begin{figure}
\includegraphics[width=\textwidth,trim=2 2 0 2,clip]{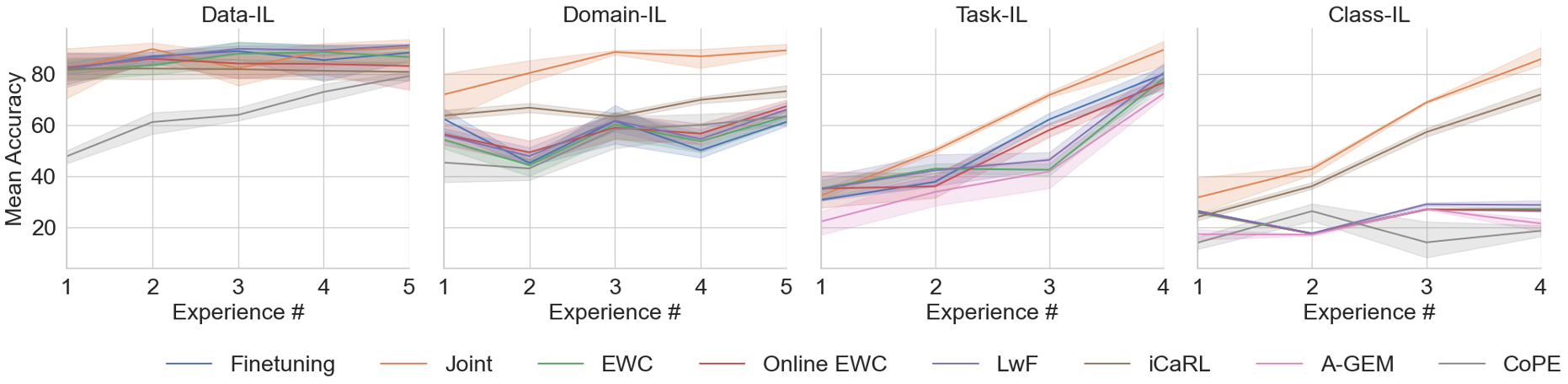}
\caption{Results from online and offline CL methods at each experience across scenarios with offline baselines.} \label{fig2}
\end{figure}
 Of the offline CL methods tested, LwF and iCaRL performed much better than finetuning in all scenarios except for Data-IL. Accuracy of EWC/online EWC was comparable to finetuning in all scenarios except for Domain-IL, where online EWC was 6.09\% higher than EWC. It should be noted that the computation time for joint baseline was much longer (250 ± 18 minutes) compared to that of for CL methods across scenarios (70 ± 8 minutes). In \textbf{Data-IL}, though CL methods were expected to outperform finetuning, we found that only LwF (accuracy: 91.32\%) outperformed finetuning and even 2.87\% higher than the upper bound joint training. While iCaRL was the best method for other scenarios, it was 7.52\% lower than finetuning. iCaRL was designed and tested for Class-IL in \cite{rksl17} and the replay memory was divided equally to store examples from each class. In our Data-IL setting, the allocated memory for each class was split further for each of the five domains. We hypothesized a bigger memory with more examples for each class per domain could have produced better results. In \textbf{Domain-IL}, iCaRL achieved the best accuracy of 73.41\%. An interesting observation is that while other methods show a reduction in performance after learning Domain 2 and 4, the performance of iCaRL continued to improve (Fig. 2). Domain 1 (original stain-normalized images) and Domain 4 (change of hue and saturation) showed minimal forgetting (Fig. S2), while Domains 2 and 3 with eosin intensity changes were much harder for CL, suggesting that intensity changes in practice due to, for example, scanner differences, may lead to larger challenges for CL than hue/saturation changes due to, for example, changes in staining reagent/condition. These 2 domains were the farthest from baseline DP settings in appearance. Note that here the 5 domains were designed to have relatively large domain gaps with non-overlapping augmentation settings and in practice by performing stain normalization during model training or for datasets with smaller domain shifts, the domain gaps would be reduced and thus potentially improving CL performance. In \textbf{Task-IL}, finetuning accuracy was much higher than that of Class-IL, likely due to the presence of task-specific network modules and additional information leveraged (taks IDs) during inference. However, unlike \cite{phh21, vfa21, vt19} have reported, LwF was the only method that performed better than finetuning (Fig. 1 Right and S3). This could be because in our network design task identities were only used in the output layer, while it was reported to be more effective to embed task identities in hidden layers \cite{de2021continual}. \textbf{Class-IL} is the most difficult learning scenario and as expected most methods suffered from severe forgetting of previous knowledge. iCaRL had the best overall accuracy of 72.06\% when tested with a random ordering of classes (Fig. S4). This is consistent with \cite{vt19} that replay methods are more effective in Class-IL than regularization methods. We also studied the impact of class order, class grouping and number of classes with iCaRL. The results (Table S2) were consistent with the hypothesis of curriculum learning that knowledge is better captured if harder tasks follow easier tasks \cite{blcw09}.\newline
\textbf{Comparison of Online CL Methods on Augmented CRC} A-GEM outperformed the online joint baseline for A-GEM (Fig. 3 Left) in Task-IL by 6\%, but suffered from catastrophic forgetting without task identity. CoPE was worse than finetuning in Data-IL and Class-IL, but was 6\% better in Domain-IL (Fig. 3 Right), demonstrating that online CL poses challenges, which calls for more attention in research. Despite lower performance compared to offline methods, knowledge retention (backward transfer) with CoPE was much better in Data-IL and Domain-IL (Fig. S1 and S2). We note that CoPE was sensitive to softmax temperature and further finetuning could help improve forward transfer. \newline
\begin{figure}[!tbp]
\includegraphics[width=1.\textwidth]{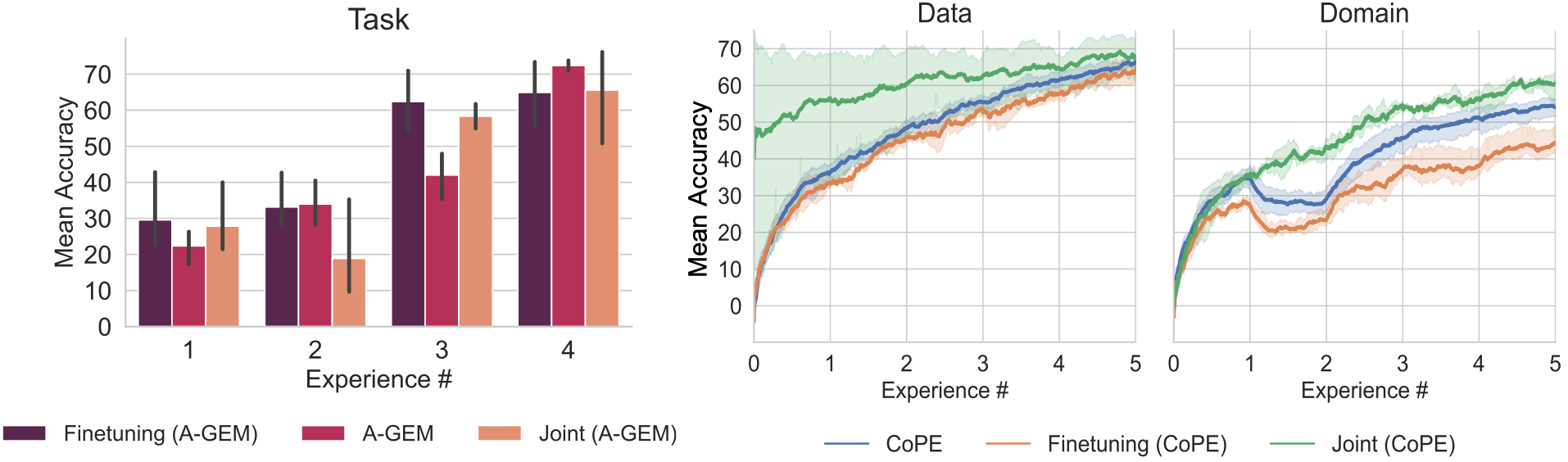}
 \caption{Left: A-GEM Task-IL accuracy at the end of each experience. Right: CoPE accuracy at each mini experience in Data- and Domain-IL with respective baselines.} \label{fig3}
\end{figure}
\textbf{Continually Learning from Multiple Tumor Types} We tested a dramatic domain shift in data streams by training a model with the original CRC images in one experience and PatchCam dataset in another experience. We experimented with LwF for continually learning the two datasets in a Domain-IL setting, where each tumor type is considered as one domain. We experimented with ordering of the domains as well as the volume of PatchCam data used in training. As shown in Fig. 4, joint baseline accuracy was around 80\%. Interestingly, LwF performed better than finetuning when PatchCam was introduced first during training followed by CRC in the subsequent experience. As expected, including 3 times more data in training improved accuracy by over 11\% (Table S2).\\
\begin{figure}
\includegraphics[width=1\textwidth,trim=2 2 2 2,clip]{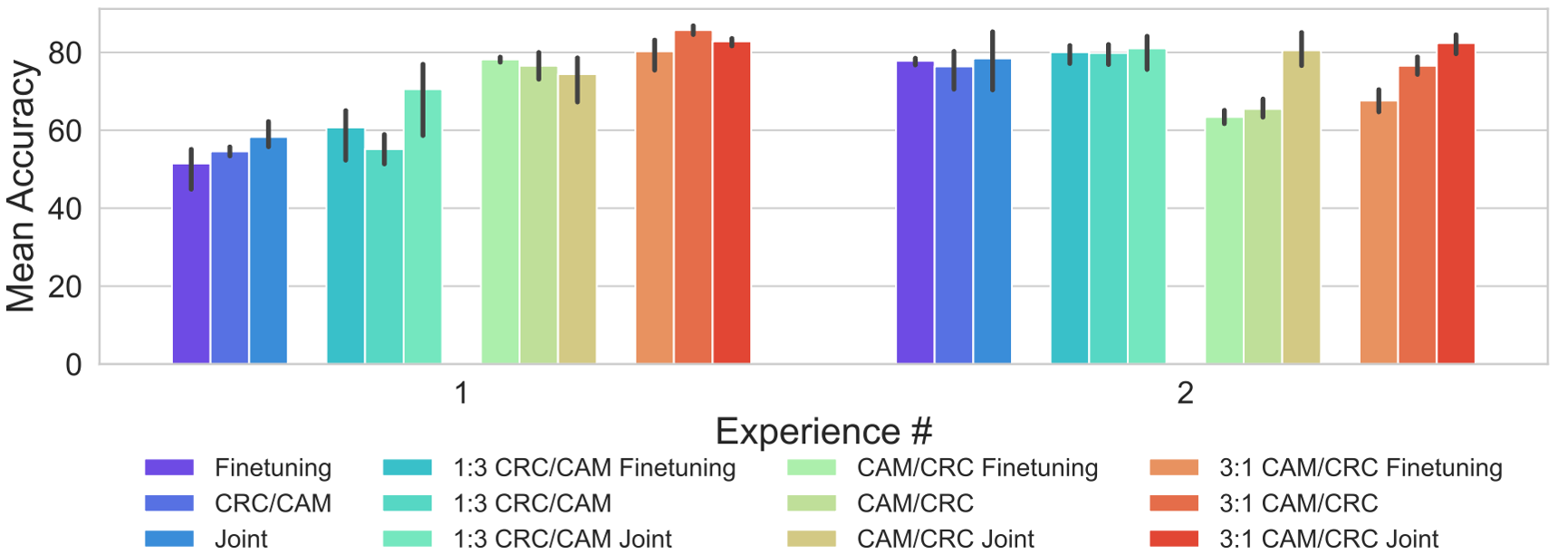}
\caption{Continual learning with CRC and PatchCam in Domain-IL} \label{fig4}
\end{figure}

\subsection{Discussion} We present a systematic study that characterizes the performance of recently proposed CL methods for different scenarios using augmented DP images with domain shifts. We found that though regularization-based methods performed  well in Data-IL and Domain-IL, only iCaRL, a rehearsal method, is effective in the most challenging Class-IL. Surprisingly, Task-IL scenarios may not be as easy to learn for DP as for other domains. CL methods were also computationally efficient, taking only about 28\% of the runtime as joint training. Though patient data evolve quickly nowadays, FDA has not approved algorithms based on CL \cite{vfa21} and extensive research is needed to establish regulations for safely incorporating CL in clinical settings. Our evaluation approaches and proposed method to generate domain-shifted datasets can potentially serve as the first step towards this goal.

\bibliographystyle{splncs04}
\bibliography{paper23.bib}

\begin{thebibliography}{10}
\providecommand{\url}[1]{\texttt{#1}}
\providecommand{\urlprefix}{URL }
\providecommand{\doi}[1]{https://doi.org/#1}

\bibitem{art18}
Aljundi, R., Rohrbach, M., Tuytelaars, T.: Selfless sequential learning. arXiv
  preprint arXiv:1806.05421  (2018)

\bibitem{ardon2021digital}
Ardon, O., Reuter, V.E., Hameed, M., Corsale, L., Manzo, A., Sirintrapun, S.J.,
  Ntiamoah, P., Stamelos, E., Schueffler, P.J., England, C., et~al.: Digital
  pathology operations at an nyc tertiary cancer center during the first 4
  months of {COVID}-19 pandemic response. Academic Pathology  \textbf{8},
  23742895211010276 (2021)

\bibitem{bgk18}
Baweja, C., Glocker, B., Kamnitsas, K.: Towards continual learning in medical
  imaging. arXiv preprint arXiv:1811.02496  (2018)

\bibitem{bayasi2021culprit}
Bayasi, N., Hamarneh, G., Garbi, R.: {Culprit-Prune-Net}: Efficient continual
  sequential multi-domain learning with application to skin lesion
  classification. In: International Conference on Medical Image Computing and
  Computer-Assisted Intervention. pp. 165--175. Springer (2021)

\bibitem{b17}
Bejnordi, B.E., Veta, M., Van~Diest, P.J., Van~Ginneken, B., Karssemeijer, N.,
  Litjens, G., Van Der~Laak, J.A., Hermsen, M., Manson, Q.F., Balkenhol, M.,
  et~al.: Diagnostic assessment of deep learning algorithms for detection of
  lymph node metastases in women with breast cancer. Jama  \textbf{318}(22),
  2199--2210 (2017)

\bibitem{blcw09}
Bengio, Y., Louradour, J., Collobert, R., Weston, J.: Curriculum learning. In:
  Proceedings of the 26th annual international conference on machine learning.
  pp. 41--48 (2009)

\bibitem{campanella2019clinical}
Campanella, G., Hanna, M.G., Geneslaw, L., Miraflor, A., Werneck Krauss~Silva,
  V., Busam, K.J., Brogi, E., Reuter, V.E., Klimstra, D.S., Fuchs, T.J.:
  Clinical-grade computational pathology using weakly supervised deep learning
  on whole slide images. Nature medicine  \textbf{25}(8),  1301--1309 (2019)

\bibitem{crre18}
Chaudhry, A., Ranzato, M., Rohrbach, M., Elhoseiny, M.: Efficient lifelong
  learning with {A-GEM}. arXiv preprint arXiv:1812.00420  (2018)

\bibitem{cruz2014automatic}
Cruz-Roa, A., Basavanhally, A., Gonz{\'a}lez, F., Gilmore, H., Feldman, M.,
  Ganesan, S., Shih, N., Tomaszewski, J., Madabhushi, A.: Automatic detection
  of invasive ductal carcinoma in whole slide images with convolutional neural
  networks. In: Medical Imaging 2014: Digital Pathology. vol.~9041, p. 904103.
  SPIE (2014)

\bibitem{de2021continual}
De~Lange, M., Tuytelaars, T.: Continual prototype evolution: Learning online
  from non-stationary data streams. In: Proceedings of the IEEE/CVF
  International Conference on Computer Vision. pp. 8250--8259 (2021)

\bibitem{he2016deep}
He, K., Zhang, X., Ren, S., Sun, J.: Deep residual learning for image
  recognition. In: Proceedings of the IEEE conference on computer vision and
  pattern recognition. pp. 770--778 (2016)

\bibitem{jm16}
Janowczyk, A., Madabhushi, A.: Deep learning for digital pathology image
  analysis: A comprehensive tutorial with selected use cases. Journal of
  pathology informatics  \textbf{7}(1), ~29 (2016)

\bibitem{khm18}
Kather, J.N., Krisam, J., Charoentong, P., Luedde, T., Herpel, E., Weis, C.A.,
  Gaiser, T., Marx, A., Valous, N.A., Ferber, D., et~al.: Predicting survival
  from colorectal cancer histology slides using deep learning: A retrospective
  multicenter study. PLoS medicine  \textbf{16}(1),  e1002730 (2019)

\bibitem{kprvdrmqra17}
Kirkpatrick, J., Pascanu, R., Rabinowitz, N., Veness, J., Desjardins, G., Rusu,
  A.A., Milan, K., Quan, J., Ramalho, T., Grabska-Barwinska, A., et~al.:
  Overcoming catastrophic forgetting in neural networks. Proceedings of the
  national academy of sciences  \textbf{114}(13),  3521--3526 (2017)

\bibitem{lss20}
Lenga, M., Schulz, H., Saalbach, A.: Continual learning for domain adaptation
  in chest {X}-ray classification. In: Medical Imaging with Deep Learning. pp.
  413--423. PMLR (2020)

\bibitem{lh16}
Li, Z., Hoiem, D.: Learning without forgetting. IEEE transactions on pattern
  analysis and machine intelligence  \textbf{40}(12),  2935--2947 (2017)

\bibitem{lr17}
Lopez-Paz, D., Ranzato, M.: Gradient episodic memory for continual learning.
  Advances in neural information processing systems  \textbf{30} (2017)

\bibitem{mna09}
Macenko, M., Niethammer, M., Marron, J.S., Borland, D., Woosley, J.T., Guan,
  X., Schmitt, C., Thomas, N.E.: A method for normalizing histology slides for
  quantitative analysis. In: 2009 IEEE international symposium on biomedical
  imaging: from nano to macro. pp. 1107--1110. IEEE (2009)

\bibitem{phh21}
Perkonigg, M., Hofmanninger, J., Herold, C.J., Brink, J.A., Pianykh, O.,
  Prosch, H., Langs, G.: Dynamic memory to alleviate catastrophic forgetting in
  continual learning with medical imaging. Nature Communications
  \textbf{12}(1),  1--12 (2021)

\bibitem{rksl17}
Rebuffi, S.A., Kolesnikov, A., Sperl, G., Lampert, C.H.: {iCaRL}: Incremental
  classifier and representation learning. In: Proceedings of the IEEE
  conference on Computer Vision and Pattern Recognition. pp. 2001--2010 (2017)

\bibitem{ruifrok2001quantification}
Ruifrok~C., J.A.: Quantification of histochemical staining by color
  deconvolution. Analytical and quantitative cytology and histology
  \textbf{23}(4),  291--299 (2001)

\bibitem{rrdskkph16}
Rusu, A.A., Rabinowitz, N.C., Desjardins, G., Soyer, H., Kirkpatrick, J.,
  Kavukcuoglu, K., Pascanu, R., Hadsell, R.: Progressive neural networks. arXiv
  preprint arXiv:1606.04671  (2016)

\bibitem{schwarz2018progress}
Schwarz, J., Czarnecki, W., Luketina, J., Grabska-Barwinska, A., Teh, Y.W.,
  Pascanu, R., Hadsell, R.: Progress \& compress: A scalable framework for
  continual learning. In: International Conference on Machine Learning. pp.
  4528--4537. PMLR (2018)

\bibitem{asmk18}
Serra, J., Suris, D., Miron, M., Karatzoglou, A.: Overcoming catastrophic
  forgetting with hard attention to the task. In: International Conference on
  Machine Learning. pp. 4548--4557. PMLR (2018)

\bibitem{slkk17}
Shin, H., Lee, J.K., Kim, J., Kim, J.: Continual learning with deep generative
  replay. Advances in neural information processing systems  \textbf{30} (2017)

\bibitem{s18}
Sornapudi, S., Stanley, R.J., Stoecker, W.V., Almubarak, H., Long, R., Antani,
  S., Thoma, G., Zuna, R., Frazier, S.R.: Deep learning nuclei detection in
  digitized histology images by superpixels. Journal of pathology informatics
  \textbf{9}(1), ~5 (2018)

\bibitem{vt19}
Van~de Ven, G.M., Tolias, A.S.: Three scenarios for continual learning. arXiv
  preprint arXiv:1904.07734  (2019)

\bibitem{vfa21}
Vokinger, K.N., Feuerriegel, S., Kesselheim, A.S.: Continual learning in
  medical devices: {FDA}'s action plan and beyond. The Lancet Digital Health
  \textbf{3}(6),  e337--e338 (2021)

\bibitem{yang2021continual}
Yang, Y., Cui, Z., Xu, J., Zhong, C., Wang, R., Zheng, W.S.: Continual learning
  with {B}ayesian model based on a fixed pre-trained feature extractor. In:
  International Conference on Medical Image Computing and Computer-Assisted
  Intervention. pp. 397--406. Springer (2021)

\bibitem{zenke2017continual}
Zenke, F., Poole, B., Ganguli, S.: Continual learning through synaptic
  intelligence. In: International Conference on Machine Learning. pp.
  3987--3995. PMLR (2017)

\bibitem{zhang2021comprehensive}
Zhang, J., Gu, R., Wang, G., Gu, L.: Comprehensive importance-based selective
  regularization for continual segmentation across multiple sites. In:
  International Conference on Medical Image Computing and Computer-Assisted
  Intervention. pp. 389--399. Springer (2021)

\end{thebibliography}

\newpage
\appendix

\section*{Appendix 1}

\renewcommand{\thetable}{S\arabic{table}}
\renewcommand{\thefigure}{S\arabic{figure}}  
\setcounter{figure}{0}    

\begin{table}
\centering
\caption{Mean and standard deviation of accuracy at the end of training from 3-5 repetitions of experiments with CL methods and their respective baselines. The average accuracies reported in Figure 1B.}\label{tab1}

\begin{tabular}{lllll} 
\hline
                   & Data-IL       & Domain-IL    & Task-IL       & Class-IL      \\ 
\hline
Finetuning         & 88.49 ± 3.15  & 61.42 ± 2.55 & 80.1 ± 5.51   & 26.61 ± 0.4   \\
EWC                & 86.77 ± 4.37  & 63.64 ± 3.2  & 78.41 ± 3.35  & 27.3 ± 0.63   \\
Online EWC         & 83.29 ± 10.02 & 67.51 ± 3.07 & 76.83 ± 2.64  & 26.79 ± 0.37  \\
LwF                & 91.32 ± 0.83  & 66.24 ± 3.57 & 80.59 ± 4.95  & 28.88 ± 2.08  \\
iCaRL              & 80.97 ± 4.51  & 73.41 ± 3.12 & -             & 72.06 ± 3.2   \\
Joint              & 90.48 ± 5.12  & 89.43 ± 2.15 & 89.58 ± 5.27  & 85.94 ± 4     \\ 
\hline
Finetuning (A-GEM) & -             & -            & 64.93 ± 9.13  & 19.57 ± 7.35  \\
A-GEM              & -             & -            & 72.41 ± 1.81  & 21.62 ± 3.45  \\
Joint (A-GEM)      & -             & -            & 65.58 ± 13.19 & 90 ± 2.72     \\ 
\hline
Finetuning (CoPE)  & 84.92 ± 4.41  & 60.85 ± 6.04 & -             & 23.76 ± 0     \\
CoPE               & 82.16 ± 4.63  & 66.71 ± 2.49 & -             & 19.15 ± 6.16  \\
Joint (CoPE)       & 86.5 ± 5.19   & 77.78 ± 8.82 & -             & 75.7 ± 0      \\
\hline
\end{tabular}

% \end{table}
\vspace{4mm} 
% \begin{table}
% \centering
\caption{(left) Impact of class order, class grouping and number of classes on iCaRL result in class-IL setting. (right) Results with LwF in domain-IL setting with CRC and PatchCam datasets with settings for domain ordering and data volume.}
\begin{tabular}{llllll} 
\cline{1-3}\cline{5-6}
\multirow{4}{*}{\begin{tabular}[c]{@{}l@{}}Class\\Order\end{tabular}}        & \begin{tabular}[c]{@{}l@{}}default \\(123456789)\end{tabular} & 78.24 ± 4.11 &         & Finetuning              & 77.84 ± 0.94  \\
                                                                             & \begin{tabular}[c]{@{}l@{}}manual \\(182736945)\end{tabular}  & 81.11 ± 3.6  &         & CRC/CAM                 & 76.39 ± 6.34  \\
                                                                             & \begin{tabular}[c]{@{}l@{}}random1 \\(794631852)\end{tabular} & 68.78 ± 7.59 &         & Joint                   & 78.44 ± 7.54  \\ 
\cline{5-6}
                                                                             & \begin{tabular}[c]{@{}l@{}}random2\\~(674398125)\end{tabular} & 68.57 ± 5.79 &         & 1:3 CRC/CAM Finetuning & 80.04 ± 2.52  \\ 
\cline{1-3}
\multirow{3}{*}{\begin{tabular}[c]{@{}l@{}}Class \\Grouping\end{tabular}}    & grouping:2223                                                 & 72.06 ± 3.2  & ~ ~ ~ ~ & 1:3 CRC/CAM             & 79.84 ± 3.34  \\
                                                                             & grouping:3222                                                 & 56.89 ± 4.69 &         & 1:3 CRC/CAM Joint       & 81.02 ± 4.7   \\ 
\cline{5-6}
                                                                             & grouping:333                                                  & 46.55 ± 1.89 &         & CAM/CRC Finetuning     & 63.44 ± 1.71  \\ 
\cline{1-3}
\multirow{2}{*}{\begin{tabular}[c]{@{}l@{}}Number \\of Classes\end{tabular}} & 9 classes                                                     & 46.55 ± 1.89 &         & CAM/CRC                 & 65.48 ± 2.95  \\
                                                                             & 6 classes                                                     & 76.69 ± 2.01 &         & CAM/CRC Joint           & 80.54 ± 4.27  \\ 
\cline{1-3}\cline{5-6}
                                                                             &                                                               &              &         & 3:1 CAM/CRC Finetuning & 67.62 ± 4.78  \\
                                                                             &                                                               &              &         & 3:1 CAM/CRC             & 76.57 ± 3.11  \\
                                                                             &                                                               &              &         & 3:1 CAM/CRC Joint       & 82.38 ± 2.5   \\
\cline{5-6}
\end{tabular}
\end{table}
\begin{figure}
%\includegraphics[width=\textwidth]{fig1.eps}
%\hspace*{-0.8cm}
\includegraphics[width=1.0\textwidth]{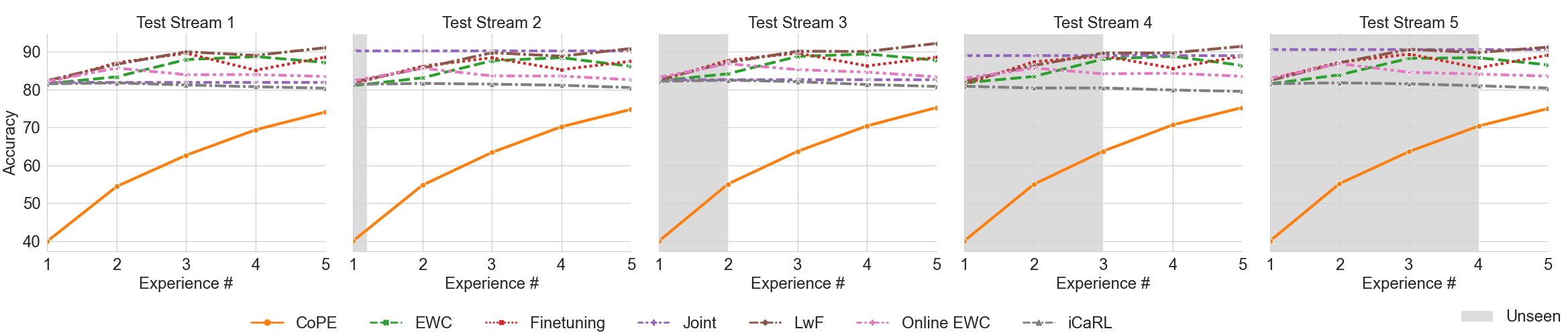}
\caption{CL method performance in Data-IL. Each subpanel shows evolution of test accuracy for a specific experience (for example, the leftmost panel shows accuracy for the test set of Experience 1, referred to as Test Stream 1, from the first experience) as more experiences are added during training. The performance of all offline CL method were comparable.} \label{figs1}
% \end{figure}
\vspace{5mm} 

% \begin{figure}
%\hspace*{0cm}
%\includegraphics[width=\textwidth]{fig1.eps}
\includegraphics[width=0.98\textwidth]{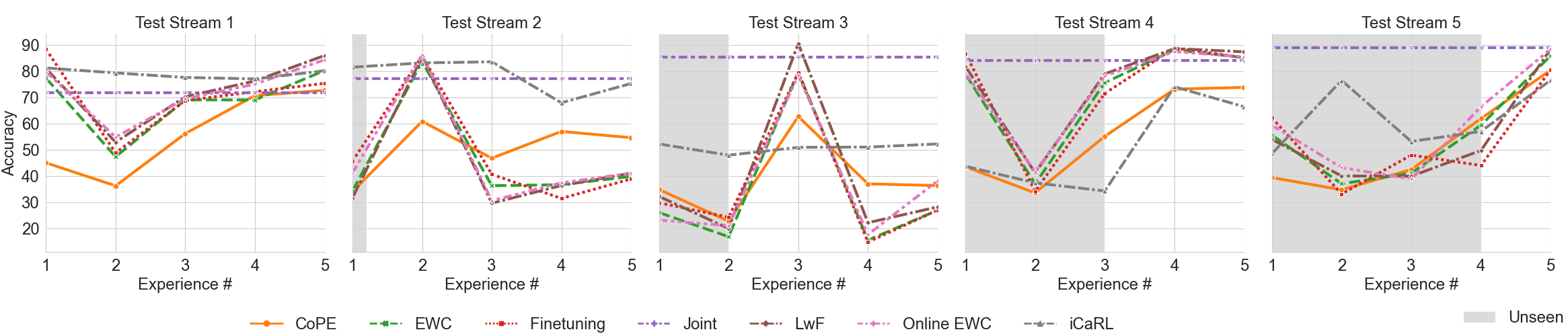}
\caption{CL method performance in Domain-IL. iCaRL achieves best accuracy among offline methods. Though CoPE performs worse than offline methods, it can be seen that it doesn't suffer from as much forgetting as other methods.} \label{figs2}
%\end{figure}

\vspace{5mm} 

%\begin{figure}
%\hspace*{-0.8cm}
%\includegraphics[width=\textwidth]{fig1.eps}
\includegraphics[width=1\textwidth]{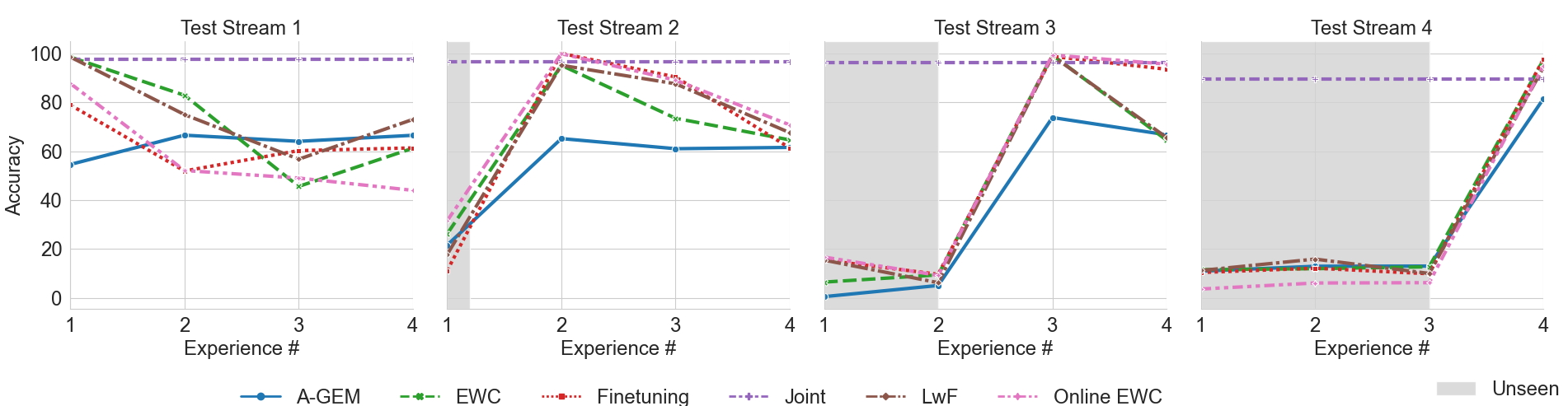}
\caption{CL method performance in Task-IL. Only LwF  outperformed fine-tuning.} \label{figs3}
%\end{figure}

\vspace{5mm} 
%\hspace*{-0.8cm}
%\begin{figure}
%\includegraphics[width=\textwidth]{fig1.eps}
\includegraphics[width=1\textwidth]{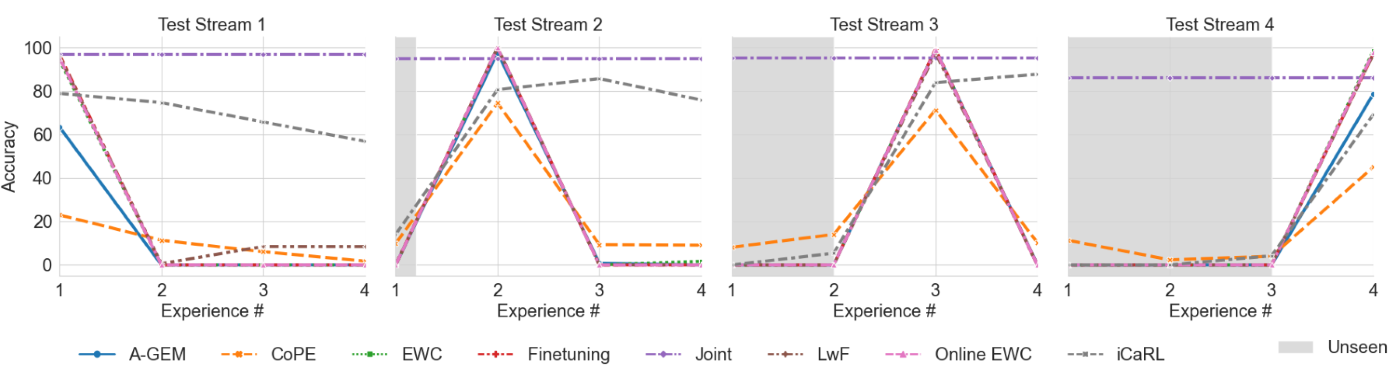}
\caption{CL method performance in Class-IL. iCaRL is the only method which does not suffer from catastrophic forgetting and performs better than fine-tuning.} \label{figs4}
\end{figure}

\end{document}